\documentclass[a4paper]{jpconf}
\usepackage{epsfig,bm,amssymb}
\begin{document}
\title{Nonlinear Kinetic Dynamics of Magnetized Weibel Instability }

\author{L. Palodhi, F. Califano, F. Pegoraro}

\address{Dept. of Phys., University of Pisa, 56127  Pisa, Italy}

\ead{lopamudra@df.unipi.it}

\begin{abstract}
Kinetic numerical simulations of the evolution of the Weibel instability during the full nonlinear regime are presented. The formation of strong distortions in the
electron distribution function resulting in formation of strong peaks in it and their influence on the resulting electrostatic waves are shown.
\end{abstract}


\section{Introduction}

The process of magnetic field generation in a plasma is one of the most important problems for both laboratory and astrophysical plasmas. Several mechanisms of magnetic
field generation have been analyzed that are effective in different plasma regimes. At high frequencies, i.e., on short electron time scales over which the ions can be
assumed to be immobile and collisions are unimportant, the Weibel instability can efficiently generate magnetic fields in plasmas with an anisotropic electron temperature distribution with zero real frequencies.

\indent The Weibel instability is an electromagnetic plasma instability driven by the presence of temperature or electron momentum anisotropy~\cite{Weibel}.

\indent In this paper we discuss the long time, kinetic, nonlinear evolution of the Weibel instability in an electron-ion plasma by using Vlasov-Maxwell simulations
in the 1D-2V phase space. Relativistic effects on the plasma dynamics will be not considered, which become significant only for very high beam speeds.

\indent The Weibel instability developed due to the counter streaming of electrons have been studied and showed that other than the electromagnetic instability, there is
also an electrostatic limit that corresponds to the well known two-stream instabilites. For more results we refer to Refs.~\cite{califano1,califano2}.  In this paper instead, we assumed temperature anisotropy to study Weibel instability  and will show, even in this case too, some kind of electrostatic instability develops due to acceleration of the particles.

\indent The Weibel instability emerges from a wide white noise spectrum of wavelengths around the electron skin depth,$d_{e}$. This length plays the role of the natural  scale length for the instability and will be used in the following as the normalizing length.

\indent In general, since the growth rate of the Weibel instability is comparable to the electron plasma frequency, and also the electromagnetic nature of the instability,
this study of the Weibel instability in the nonrelativistic limit has been limited so far to the electron dynamics. This approximation, which considers the ions as fixed neutralizing background, is valid during the linear and very early nonlinear phase of the instability. On the other hand, for longer times, the ion response cannot be neglected.     

\indent The paper is organized as follows. In the next section we will introduce the governing equations and the initial conditions. The numerical simulations are presented
in section III. Conclusions follow in section IV.

\section{The Equations}

The kinetic evolution of an electron-ion collisionless plasma is  described by the Vlasov equation in the phase space, coupled self-consistently to the Maxwell equations.
We have used 1D-2V $(x,v_{x},v_{y})$ phase space, with periodic boundary condition along $x$ direction, unidirectional magnetic field $B_{z}$ and a two-component electric
field $\mathbf{E} = (E_{x},E_{y})$.

\indent The Vlasov-Maxwell equations are normalized by the following characteristic quantities: the electron mass $m_{e}$, the characteristic electron density $\bar{n_{e}}$, the speed of light c, the electron plasma frequency $\omega_{pe} = ( 4\pi\bar{n_{e}}e^2 / m_{e} )$, the electron skin depth $d_{e} = c / \omega_{pe}$ and the characteristic  fields $\bar{E} = \bar{B} = m_{e}\omega_{pe}c/e$. In  dimensionless limits the Vlasov-Maxwell equations
for electrons read as

\begin{eqnarray}
\frac{\partial{f}_{e}}{\partial{t}} + \mathbf{v}.\frac{\partial{f}_{e}}{\partial{\mathbf{x}}} + (\mathbf{E} + \mathbf{v} \times \mathbf{B}) \cdot\frac{\partial{f}_{e}}{\partial{\mathbf{v}}} = 0
\end{eqnarray}

\begin{eqnarray}
\frac{\partial{B_{z}}}{\partial{t}} = -\frac{\partial{E_{y}}}{\partial{x}}, \hspace {4mm} \ \frac{\partial{E_{y}}}{\partial{t}} = -\frac{\partial{B_{z}}}{\partial{x}} + J_{y}
\end{eqnarray}
where $f_{e}(x,v_{x},v_{y},t)$ is the 1D-2V electron distribution function (edf). The Poisson equation given by  $\frac{\partial^2\phi}{\partial^2x} = (n_{i}-n_{e}) $, where, $\phi$ is the potential and $n_{i}$ and $n_{e}$ are ion and electron density respectively.

\indent We integrate Vlasov equation in the interval $L_{x} = 2\pi/k$  to have the total length of the box large enough to include the most unstable mode in the simulation. The initial dimensionless distribution function considered is:

\begin{eqnarray}
f_{e}(x,v_{x},v_{y},0) = \Big(\frac{n_{e}}{\pi v_{th}^2}\sqrt{\frac{T_{y}}{T_{x}}}\Big)exp\Big[\frac{1}{v_{th}^2}\big(-v_{x}^2-\frac{T_{x}}{T_{y}}v_{y}^2\big)\Big][1 + \epsilon \sum^{300}_{n=1}\ \cos(kx+\phi(k))]
\end{eqnarray}
where, $L_{x}$ is the length in the $x$ direction, $v_{th}$ is  the thermal velocity of the electrons and  $T_{x}$ and $T_{y}$ are the temperatures along $x$ and $y$ direction 
respectively, the modes of the wave, $k=2\pi n/L_{x}$, and $\epsilon$ is the perturbation amplitude. To include all the modes possible we choose 300 as the summation limit. At $t=0$ we introduce also a  perturbation on the magnetic field:

\begin{eqnarray}
B_{z} = a_{in} \sum^{300}_{n=1}\ [\cos(k_{n}x+\phi(k))]
\end{eqnarray}
where, $a_{in}$ is the magnetic field amplitude and is independent of the mode.

\section{Numerical Simulations}

We have performed the numerical simulation(see Ref.~\cite{Mang} for the numerical algorithm) of the Weibel instability in  plasma, by integrating numerically Eqs.(1)-(2) with the initial conditions given by Eqs.(3)-(4). In particular, the electron and proton distribution function is advanced in time using the electromagnetic splitting method  in the 1D-2V limit. The Maxwell equation are integrated in time by using the third order Adams - Bashforth algorithm and standard Fast Fourier Transform technique are applied to calculate the space derivatives.
The temperature anisotropy considered is  $T_{y}/T_{x} = 12$.  We have performed the simulation with higher and lower temperature anisotropies also. But with lower temperature anisotropies the growth of the  magnetic field is not strong enough to see the features we are interested in and also take longer time to run the simulation to reach the nonlinear regime. Nothing interesting is found at higher temperature anisotropies too. Hence, for the sake of qualitative agreement we assumed the particular case of anisotropy with $T_{y}/T_{x} = 12$. The thermal velocity is taken as  $v_{th} = 0.02$, corresponding a plasma with a non-relativistic temperature. The perturbation  amplitude for each wave is $a_{in} = 10^{-4}$. Initially, we consider a field free plasma, hence $<B_{z}> = 0$. The number of points in the $x$ and $v$ directions are $N_{x} = 600$ and $N_{v_{x}} = 90$, $N_{v_{y}}= 90$. The time step is $\Delta t = 0.005$. The evolution of the system is investigated up to $t = 480\omega_{pe}^{-1}$. The maximal velocity which defines the velocity interval is set to $v_{x_{max}} = 0.1$ and  $v_{y_{max}} = 0.3$ in the $x$ and $y$ directions respectively such that $-0.1\le v_{x}\le 0.1$ and $-0.3\le v_{y}\le 0.3$. The length of the space domain is $L_{x} = 6\pi$. The space discretization in $x$ is $dx = 0.03d_{e}$, where $d_{e}$ is the electron skin depth.

\indent In Fig.1 (A) we plot the Fourier amplitudes of the magnetic field $B_{z,k}$ vs $k$ at $t=225$. We see the modes in the interval $0.1\le k \le 3$ have grown with a maximum growth rate corresponding to $k=1.25$. Note that a small portion of interval in $k$ is taken. In (B) we plot the Fourier amplitude of the magnetic field and of the $x$ and
$y$ components of the electric field. In the linear phase, $t>100$, the magnetic field grows exponentially, corresponding to a growth rate $\gamma=0.0390$. During this phase
the Fourier amplitudes of the electrostatic field $E_{x,k}$ remains negligible. However from  $t=190$ to  $t<300$, the electrostatic field starts to grow at a very rapid rate, of the order of twice the linear growth rate of the instability, $\gamma_{es}=2\gamma$($\gamma_{es}$ is the electrostatic growth rate),  a clear signature of the beginning of the non linear regime. At later time, $t>300$ the magnetic field saturates. The electrostatic field levels off at $t>300$ due to particle trapping in phase space.

\begin{figure}[!t]
\centerline{\psfig{figure=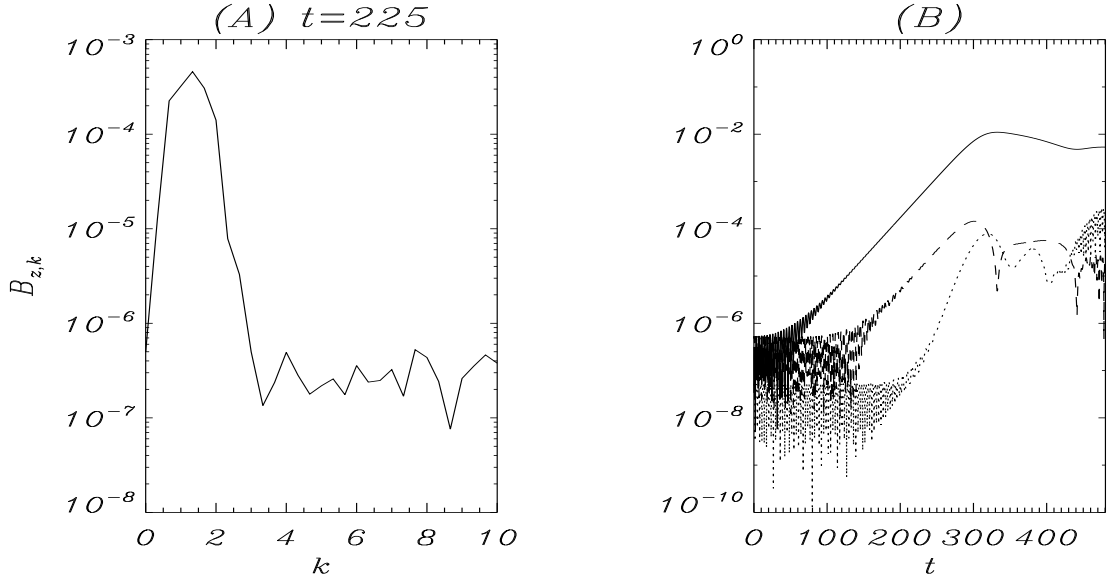,height=8cm,width=16cm}}
  \caption[ ]{\small Fig(A) shows the Fourier transform of the magnetic field $B_{z,k}$ vs $k$ at $t=225$ showing the most unstable mode at $k=1.25$. Fig(B) plots the time
evolution of  $B_{z,k}$(solid lines), $E_{x,k}$(doted lines), $E_{y,k}$(dashed lines) for  $k=1.25$.
}
\label{Fig6}
\end{figure}

\begin{figure}[!t]
\centerline{\psfig{figure=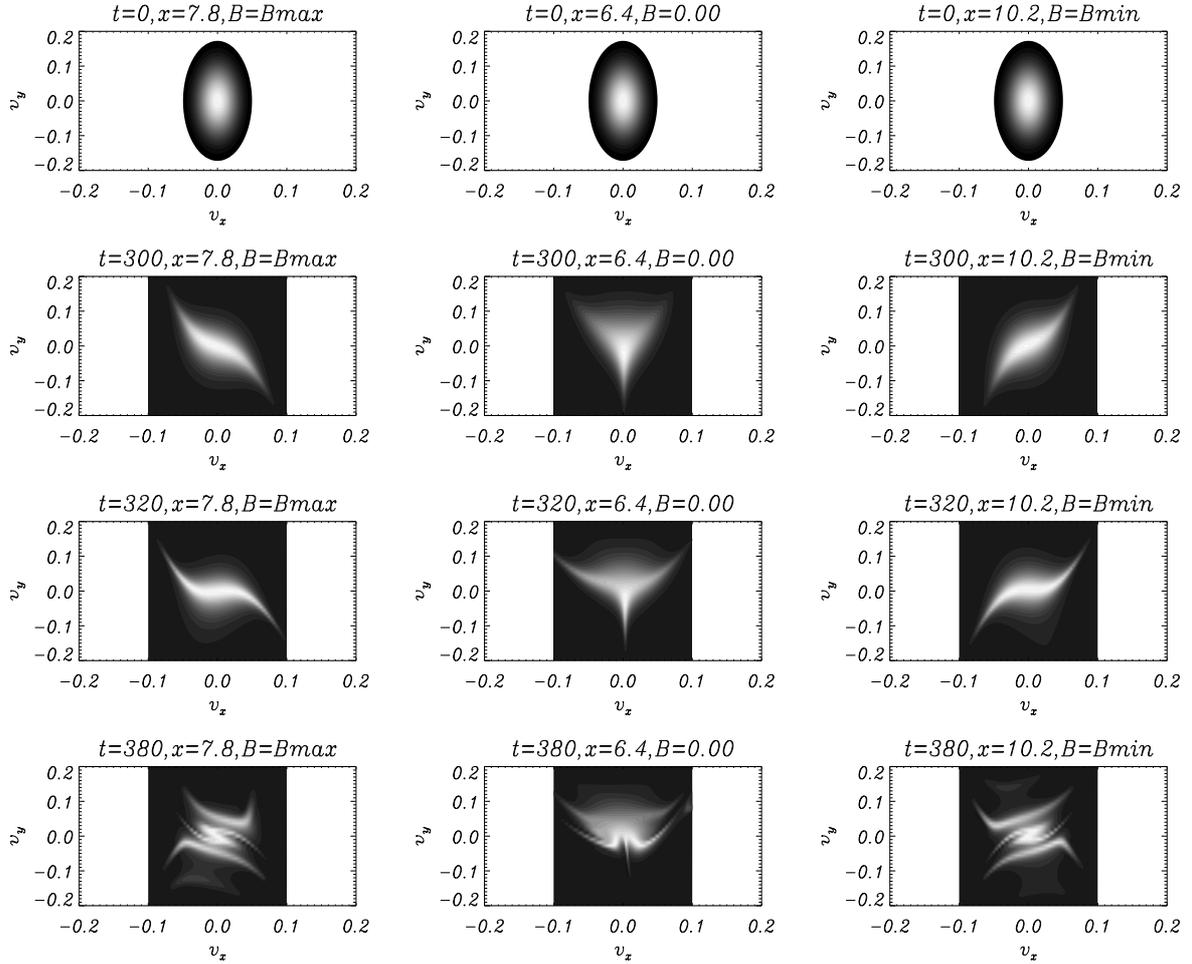,height=13cm,width=16cm}}
  \caption[ ]{ Contour plot of the distribution function at $t = 0,300,320,380$ at $x = 7.8, 6.4, 10.2$  which are the points of maximum, zero and  minimum amplitude of the magnetic field.
}
\label{Fig6}
\end{figure}

\begin{figure}[!t]
\centerline{\psfig{figure=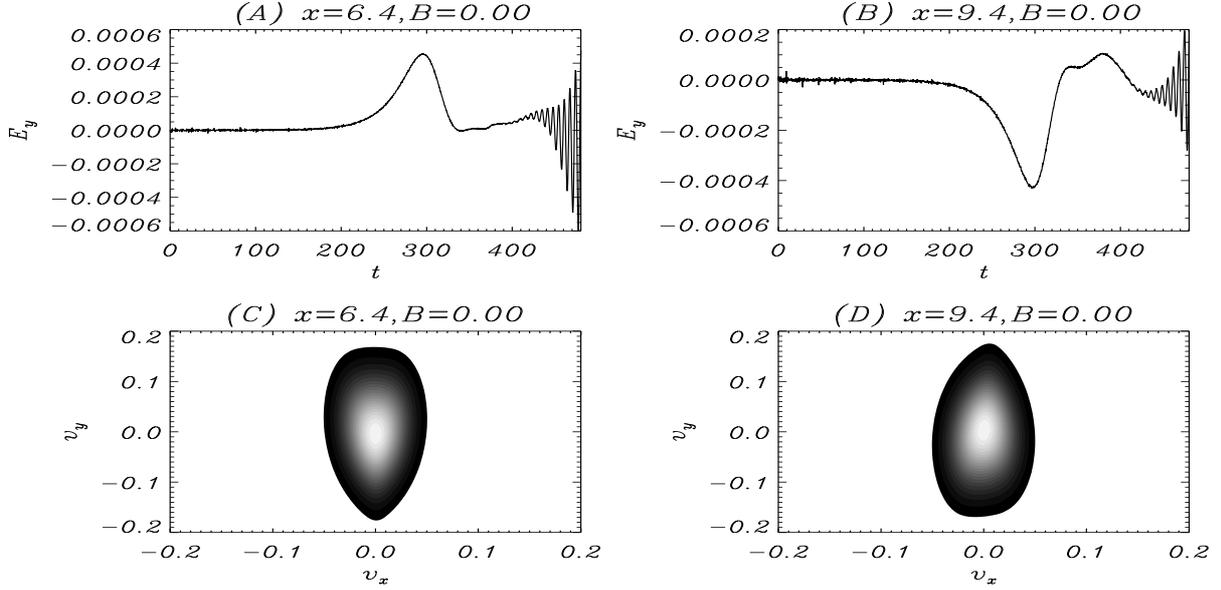,height=8cm,width=16cm}}
  \caption[ ]{ Frames
(A, B) plot induced electric field $E_{y}$ vs time at $x= 6.4, 9.4$ where $B_{z} = 0$. Frames (C, D) plot contour of edf in $(v_{x},v_{y})$ phase plane at  the same positions.
}
\label{Fig6}
\end{figure}


\indent In Fig.2 we plot the electron distribution function(edf) in $(v_{x},v_{y})$ space at time $t = 0,300, 320, 380$. Frames in the first row show the edf at $t=0$ at
the position where the magnetic field is maximum, zero and minimum. Next rows show the same at different times.  At  $t = 0$  we can see that the velocity in the $y$ direction
of the particles is greater than in the $x$ direction, corresponding to our initial condition where, in the $y$ direction the particles are hotter in the $x$ direction. A deformation  of the edf is observed from  $t>300$. We see that there is a $x$-dependent rotation of the edf  at the positions of maximum and minimum magnetic  field at each time. On the contrary there is net acceleration of electrons in the  direction of the electrostatic force at the position of zero magnetic field. We also observe that with increase in time additional structures in velocity space are formed elongated along $v_{x}$. These additional structures in phase space may correspond to the particles that have been accelerated by the induced electric field and deflected by the  magnetic field.  To see further, we plot in Fig.3 in  frames (A) and (B), the induced electric field vs time at  $x = \pi/2k$ and $x=3\pi/2k$, positions corresponding to first and second zero's in the magnetic field. We see that the induced electric field at first zero  is positive while for the next it is negative, which relates to the acceleration of the particles as mentioned earlier. Same kind of distortions are observed in the corresponding contour plots in (C) and (D).

\begin{figure}[!t]
\centerline{\psfig{figure=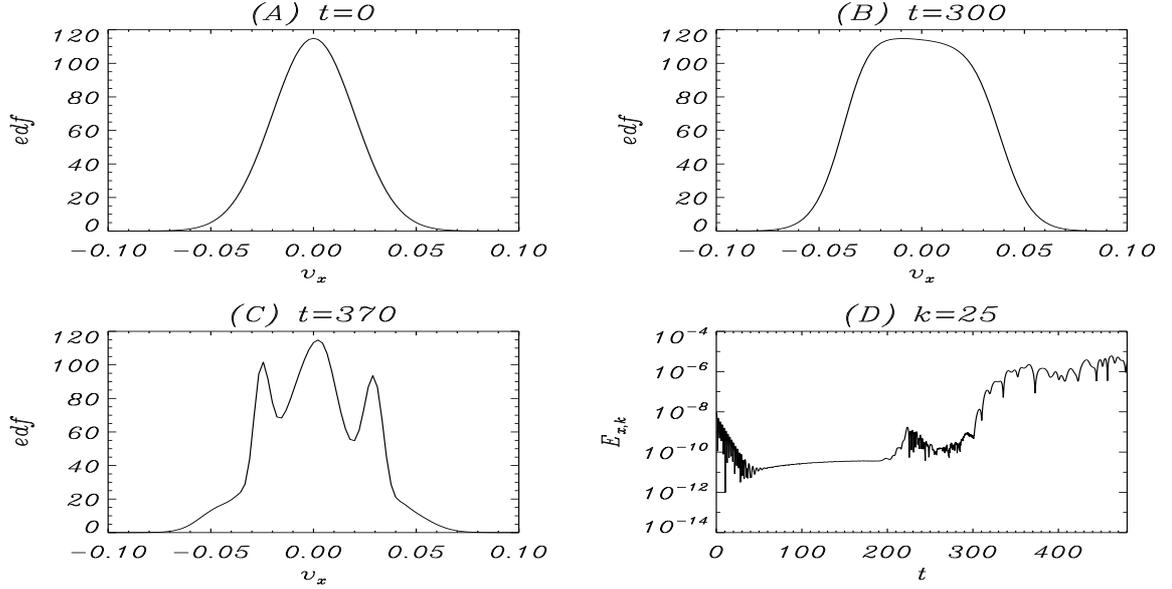,height=8cm,width=16cm}}
  \caption[ ]{\small Frames (A,B,C) show the time evolution of  edf versus $v_{x}$ at $t=0,300,370$. Frame (D) plots the time evolution of $E_{x,k}$ at $k=25$.
}
\label{Fig7}
\end{figure}

\indent As shown earlier, for $t<300$ the electrostatic field grows exponentially. In Fig.4, frames (A)-(B)-(C) show the time evolution of the electron  distribution function versus $v_{x}$ at fixed spatial position. We observe bumps along $v_{x}$ direction of the distribution function at $t\ge300$. These bumps remain  for quite a  long time up to
$t=370$ with a resonant velocity $v_{r} = 0.04$. In the last frame we show for corresponding k=25 the electrostatic field in time. We observe in the plot that indeed
the electrostatic field grows in time during the period when we observe bumps in the edf.

\section{\small Conclusions}

With the help of the numerical simulations, we have shown the kinetic saturation of the Weibel instability. Furthermore, it has been shown that the evolution of the Weibel
instability  leads to distortions in the electron distribution function that give rise to the resonant onset of electrostatic field.

\end{document}